\documentclass[showpacs,twocolumn,prl]{revtex4}
\usepackage{graphicx}
\usepackage{epsfig}
\usepackage{amssymb}

\def\beq{\begin{equation}}
\def\eeq{\end{equation}}
\def\beqa{\begin{eqnarray}}
\def\eeqa{\end{eqnarray}}

\def\e{\epsilon}

\def\half{{\ss 1\over 2}}

\def\D{\Delta}

\def\e{\varepsilon}
\def\cH{{\mathcal H}}

\def\ss{\scriptstyle}

\def\r{{\bf r}}

\def\ss{\scriptstyle}
\def\al{\alpha}
\def\be{\beta}

\def\la{\lambda}
\def\etal{{\sl et al.}}
\def\o{\omega}

\def\nonum{\nonumber \\}

\def\nonum{ \nonumber \\}

\renewcommand{\sec}[1]{\vskip 0.truecm \noindent {\sl #1}. -- }

\begin{document}

\title{Impurity induced bound states and proximity effect in a bilayer exciton condensate}
\author{ Yonatan Dubi$^1$ and Alexander V. Balatsky$^{1,2}$}
\affiliation{$^1$ Theoretical Division, Los Alamos National Laboratory, Los Alamos, NM 87545, USA}
\affiliation{$^2$
enter for Integrated Nanotechnologies, Los Alamos National Laboratory, Los Alamos, NM 87545}
\begin{abstract}
The effect of impurities which induce local interlayer tunneling in
bilayer exciton condensates is discussed. We show that
a localized single fermion bound state emerges inside the gap for any  strength of impurity scattering and calculate the
dependence of the impurity state energy and wave function on the potential strength. We show that such an impurity induced single fermion state enhances
the interlayer coherence around it, and is similar to  the
superconducting proximity effect. As a direct consequence of these single impurity states, we predict
that a finite concentration of such impurities will increase the
critical temperature for exciton condensation.   \end{abstract}
\maketitle

\sec{Introduction} The Bose condensation of electron-hole pairs (excitons) in semiconductors is an old idea \cite{Keldysh,Zhu} which received renewed
attention, mostly due to the possible experimental realization of such a condensate in semiconductor bilayers \cite{Eisenstein1,Butov1}. In these
systems, two quantum wells are separated by an insulating barrier, which prevents fast recombination of the excitons and allows for a coherent exciton condensate (EC) to develop. The lack of direct tunneling between the layers is thus a crucial component in the existence of an EC. The role of interlayer tunneling has been studied some time ago \cite{Shevchenko}, and it was shown that interlayer tunneling is not necessarily detrimental to the EC, although it may induce finite dissipation in the current flow, which may explain the failure to observe pure dissipationless current flow in these systems.

The key issue in the experimental verification of an EC is the identification of a clear signature that provides a convincing proof of EC. Indirect evidence for exciton condensation has been provided by tunneling experiments  \cite{Spielman,Champagne}, vanishing Hall resistance \cite{Kellogg}, photoluminescence \cite{Butov2,Butov3} and pattern formation \cite{Butov4} in photoexcited indirect excitons to name the few. Yet, in the absence of direct evidence of dissipationless supercurrent, it is important to devise other methods in which the properties of the EC may be probed.

In this paper we suggest that the presence of an EC may be determined by studying its response to local impurities. This notion, of studying impurities to determine the structure of an underlying condensate structure, was suggested in the context of  excitonic condensate \cite{Bistritzer} and d-wave superconductors {\cite{Sasha1, Sasha2,Fischer}, and was expanded to various systems such as bilayer cuprate superconductors \cite{Zhang}, inhomogeneous cuprates \cite{Tsuchiura}, iron-based superconductors \cite{Tsai}, exotic superconductors \cite{Zuo}, and various Graphene systems \cite{Dahal,Wehling}. While in these cases the impurities are either scattering or magnetic impurities, as we will show below the bilayer exciton system will support impurities of another kind, somewhat analogous to negative-U impurities in superconductors \cite{Taraphder,Litak,Dubi}.

Consider  a bilayer system, composed of two quantum wells placed one on top of another (with an insulating barrier in between), in which at a certain point defect  in space the two layers become close enough to allow  greater direct interlayer tunneling. This local defect is not a hole in tunneling barrier but a region of weaker tunneling gap.  We call this point the tunneling impurity (Fig.~\ref{diagram}(a)).
  \begin{figure}[ht]
\vskip 0.5truecm
\includegraphics[width=8truecm]{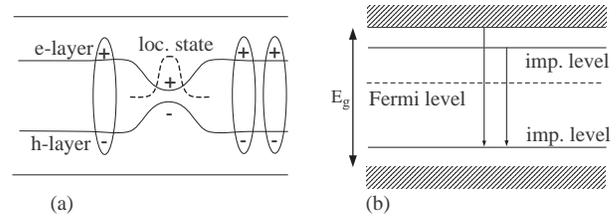}
\caption{(a) Real-space schematic representation of the local tunneling impurity. Far from the impurity electron-hole pairs form the exciton condensate, but at the tunneling impurity localized bound states are formed. (b) Energy-space representation of our results, showing the energy position of the impurity-induced bound states. New spectroscopic features will emerge as a result of these bound states, depending on their occupation. A few photoluminescence processes due to the bound states are illustrated. }\label{diagram}
\end{figure}

  Clearly, if there are too many such tunneling impurities, the excitons will recombine before the EC is achieved. Here we wish to study the case of either a single tunneling impurity or a finite (yet small) concentration of such impurities. Our main results are as follows:
\begin{itemize}
\item For a single tunneling impurity, we find that a sub-gap bound state is formed, at an energy $\o_0+\be=\pm \D \sqrt{1-\al^2} \sqrt{\frac{1-\Gamma}{1+\Gamma}}$ (Fig.~\ref{diagram}(b)), where $\al$ and $\beta$ define the band structure and $\Gamma$ is defines the impurity strength (see Eq.(\ref{X}-\ref{Imp-energy})). The spatial extent of the bound state is given by a length-scale $\xi \sim \sqrt{\frac{\Gamma}{1+\Gamma}}$. This result is non-perturbartive, and applies to any tunneling strength.
\item In the vicinity of the tunneling impurity, the interlayer coherence is \emph{enhanced}.
\item Following, a finite concentration of tunneling impurities should result in an increase of the EC critical temperature.
\end{itemize}

 We suggest to test our predictions by deliberately introducing such impurities into the bilayer systems (for instance by ion bombardment). Our results should apply to both EC formed in quantum hall bilayer and in photoexcited exciton systems. There is evidence in optical measurements of bilayer systems that such impurities are formed in the growth process \cite{Butov1}.

\sec{Single-impurity bound state}

The starting point for this calculation is the usual mean-field description of the bilayer system \cite{Zhu,Pieri,Shim}
\beq
 \cH_{MF}=\sum_{\al k}\e_{k\al }c^{\dagger}_{k\al }c_{k\al}+\D\sum_{k}\left( \D_k c^{\dagger}_{k+}c_{k-} +\text{h.c.} \right)
 \label{H_MF} \eeq
 where $+(-)$ refers to the upper (lower) layer, and $\e_{k+}=\frac{\hbar^2}{2m_+} k^2,~\e_{k-}=-\frac{\hbar^2}{2m_-} k^2-E_g$ (the chemical potentials can be absorbed into  $E_g$). The order parameter $\D_k$ should in principle be determined self consistently, but for the sake of allowing for an analytic calculation we will assume its value is known. Moreover, we will assume that it takes a similar form to that of the superconducting gap, i.e. it is finite (and uniform) within some range from the Fermi energy. This is not a bad approximation when the EC is of a BCS-like nature \cite{Pieri}. We also point that the real spin of the electrons has been disregarded, as it plays no significant role in the situation we describe here.

 The hamiltonian of Eq.~(\ref{H_MF}) is similar to the BCS hamiltonian, and it is thus useful to follow the formulation used to study single impurity states in superconductors \cite{Sasha2}. We define Nambu-like operators, $\psi_k=\left( \begin{array}{c}c_{k+} \\c_{k-} \\ \end{array}\right)$, for which the Green's function may be written as a $2 \times 2$ matrix
 \beq
 \hat{G}_{k}(t)=\langle \psi^{\dagger}_k (t) \psi_k(0) \rangle=\left(
            \begin{array}{cc}
              G_{k,+} & F_k \\
              F^\dagger_k & G_{k,-} \\
            \end{array}
          \right) ~~.\eeq
In the absence of impurities, the electron (+), hole (-) and anomalous Green's functions (in energy domain) are given by
\beqa
 g_{+}&=& \frac{\o-E_k-\e_k}{(\o-E_k)^2-\e^2_k-\D^2 },~
 g_{-}= \frac{\o-E_k+\e_k}{(\o-E_k)^2-\e^2_k-\D^2 }\nonum
 f &=& \frac{\D}{(\o-E_k)^2-\e^2_k-\D^2} ~~,\label{GreenFunction} \eeqa where $E_k=\half (\e_{k+}+\e_{k-}),~\e_k=\half (\e_{k+}-\e_{k-})$, and the explicit dependence on $k$ has been omitted for convenience. We note that, as opposed to the BCS case, the electron and hole Green's functions are not symmetric, due to the unequal masses (and hence the different band structure).

We now turn to the local tunneling impurity. In real space, one can imagine it as a point in which the layers are closer to each other, and hence tunneling there is amplified (Fig.~\ref{diagram}(a)). Thus, the impurity hamiltonian is \beq
 \cH_{imp}=-\la \psi(0)^\dagger_{+}\psi(0)_{-}+\text{h.c.} =-\la \sum_{kk'} c^\dagger_{k+}c_{k'-}+\text{h.c.} \label{H_imp} ~~\eeq The sign of $\lambda$ defines the nature of the coupling between the single-particle states of the two layers, with a positive (negative) $\lambda$ corresponding to bonding (anti-bonding). The first is the more natural situation, but one can imagine an anti-bonding situation if, for instance, the interlayer tunneling is mediated by non-s-wave orbitals in the layer separating the two quantum wells. As will be evident from the results, the sign of $\lambda$ does not have any significant effect on the final outcome.

  In the language of the Nambu operators
    the impurity hamiltonian takes the form \beq \cH_{imp}=-\la \sum_{kk'} \psi^\dagger_k \tau_1 \psi_{k'} \eeq where $\tau_i,~i=0,...,3$ are the Pauly matrices. The constant $\lambda$ describes the local tunneling amplitude between the layers.

To continue, we follow the prescription used by Shiba and others \cite{Shiba,Sasha2} and introduce the $T$-matrix, defined via the Dyson equation for the Green's function in the presence of the impurity, \beq \hat{G}=\hat{g}+\hat{g} \hat{T} \hat{g} ~~, \label{Dyson_Eq}\eeq where $\hat{g}$ is the bare (Nambu) Green's function. For a perfectly local impurity (as assumed here), the interaction vertex does not depend on momentum and is given by $\hat{U}=-\lambda \tau_1$. The $T$-matrix is determined by the equation \beq
  T(\o)=\hat{U}+\hat{U} \sum_k \hat{G}_k T(\o) \label{T-matrix_Eq}\eeq and
   \beq
  \hat{X}=\sum_k \hat{G}_k= \frac{2\pi i N_0}{ \sqrt{(\o+\be)^2-(\al^2-1)\D^2}} \left(
                                                                         \begin{array}{cc}
                                                                           \frac{\o+\be}{\al-1} & -\D \\
                                                                           \D & -\frac{\o+\be}{\al+1} \\
                                                                         \end{array}
                                                                       \right) ~~,\label{X}\eeq
  where $\al=\frac{m_- - m_+}{m_-m_+},~\be =\frac{m_+}{m_-+m_+} Eg$, and $N_0$ is the two-dimensional density of states with the reduces mass.
  It is now a matter of straight-forward algebra to evaluate the $T$-matrix. The position of the single-particle level induced by the impurity potential is determined from the position of the poles of the $T$-matrix, which are given by  \beq
  \o_0+\be=\pm \D \sqrt{1-\al^2} \sqrt{\frac{1-\Gamma}{1+\Gamma}} ,~~\Gamma=\frac{4 \pi^2 \la^2 N_0^2}{1-\al^2}~. \label{Imp-energy}\eeq
The real-space length scale associated with the impurity state may be found by evaluating the real-space dependence of the single-particle Green's function. This amounts to performing the inverse Fourier transform of the Green's function (Eq.~(\ref{Dyson_Eq}) with the help of the solution of Eq.~(\ref{T-matrix_Eq}), and we find that the real-space structure has an exponential decay around the impurity site (located at ${\bf r}=0$)  with a length scale $\xi \sim \sqrt{\frac{1+\Gamma}{\Gamma}}$. As $\Gamma \to 0$ the impurity state merges with the regular excitations, becomes a plain-wave and hence has a diverging $\xi$.

There are several reasons why these impurity-induced single-fermion bound states are important. First, since they are optically active (i.e. one can optically excite them and induce transition between them and the regular excitations), they should be in principal observable to spectroscopy experiments. Second, they point out to the fact that a simple mean-field approach to disorder in bilayer systems \cite{Bistritzer} may not be enough to adequately characterize the effect of disorder. Finally, as we show below, they induce interlayer coherence in their vicinity and thus may increase the critical temperature. In addition, since they are experimentally achievable and due to the analogy with superconducting negative-U impurities, they may shed light on the physics of the latter, which are not experimentally accessible.

\sec{Finite impurity concentration}
Next we turn to the effect of a finite impurity concentration. The usual treatment of this case dates back to Abrikosov and Gor'kov \cite{Abrikosov}, yet it involves averaging over impurity positions, and thus fails to produce the single impurity physics which we are interested in. For that reason, we choose a real-space approach, by solving the bilayer problem numerically on a square lattice. The Hamiltonian is given by
\beqa
\cH &=& -t\sum_{\langle i,j \rangle ,\al}c^\dagger_{i \al}c_{j,\al}+ \sum_{i,\al} E_{\al}c^\dagger_{i,\al}c_{i,\al}  \nonum & & +\sum_{i,j} \left( \D_{i,j} c^\dagger_{i,+}c_{j,-}+\text{h.c.} \right)-\la \sum_{j}\left( c^\dagger_{j,+} c_{j,-} \right) ~~, \nonum & &\label{Real Space H} \eeqa
where again $\al=\pm$ corresponds to the electron and hole layers, with $E_\pm=\pm E_g/2$ (the chemical potentials are absorbed into this energy, and we keep the populations equal, as well as the effective masses). $t$ is the usual tight-binding (intralayer) hopping parameter, and $t=1$ sets the energy scale hereafter. The order-parameter $\D_{i,j}$ is calculated self-consistently via
\beq
\D_{i,j}=\frac{U}{|\r_i-\r_j|} e^{-|\r_i-\r_j|/\xi} \langle c^\dagger_{i,+} c_{j,-} \rangle ~~,\label{D_self_consistency} \eeq
where $U$ is the strength of the coulomb interaction, $|\r_i-\r_j|$ is the distance between the two sites labeled $i$ and $j$ (including the interlayer separation $d$), and $\xi$ is some screening length for the Coulomb interaction, which in principal should be determined from the intralayer Coulomb screening. We have tested our results for different values of $\xi$ and found no qualitative difference between them. However, small $\xi$ allowed for better numerical convergence, and thus the results presented below were performed with $\xi=1$ (in units of lattice spacing). In the last sum of Eq.~(\ref{Real Space H}) $\la$ is the interlayer tunneling strength, and the sum is performed over a randomly chosen set of sites \{ j \} which comprises a fraction $p$ of the entire lattice.

The numerical calculations were performed until local self-consistency was achieved for both the order parameter $\D_{i,j}$ and the local density (which we kept at $n_+=n_-=0.46$, i.e. slightly below half-filling). From $\D_{i,j}$ we define a local order parameter $\D_i=\sum_j \D_{i,j}$. We have performed our numerical calculations with various parameters (i.e. lattice size, interaction strength, impurity concentration) and have found similar results in all of them.

In Fig.~\ref{Single_Impurity} we show the average local order parameter $\bar{\D}=\sum^{'}_{i} \D_i$ as a function of temperature, for different values of the tunneling amplitude $\la=0,0.1,...,0.5$. The tunneling impurity concentration is $p=0.1$ and the sum is over sites which do not have a tunneling impurity in them, which means that we are probing the influence of an impurity on its vicinity. Other numerical parameters are: system size $25 \times 25$ lattice sites, interlayer distance (in units of the lattice spacing) $d=0.5$, and interaction strength $U=1$. One clearly sees that a finite impurity concentration results in an increase in $T_c$ and an apparent smoothing of the transition. Both these effects should be observable in experiment.
In the inset of Fig.~\ref{Single_Impurity} we plot the real-space structure of the order parameter along one direction in a system with the same parameter except the presence of only one impurity (with $\la=0.2$), located at the center of the lattice, for difference temperatures, from $T=0.01$ o $T=0.3$ (with the direction of the dotted arrow). At the temperature which corresponds to $T_c$ for $\la=0$ there is a clear jump in the order parameter, yet it remains finite on sites in the vicinity of the tunneling impurity. This strongly resembles the behavior of the superconducting order-parameter in the vicinity of a negative-U impurity \cite{Dubi}, i.e. a proximity effect. Interestingly, we found from our numerical calculations that the spatial dependence of the order parameter (as a function of its distance from the impurity) is approximately given by $\D(r)\sim \exp(-\left(\frac{r}{r_0}\right)^{1/2})$, with the length-scale $r_0$ independent of the tunneling amplitude $\la$.
\begin{figure}[ht]
\vskip 0.5truecm
\includegraphics[width=8truecm]{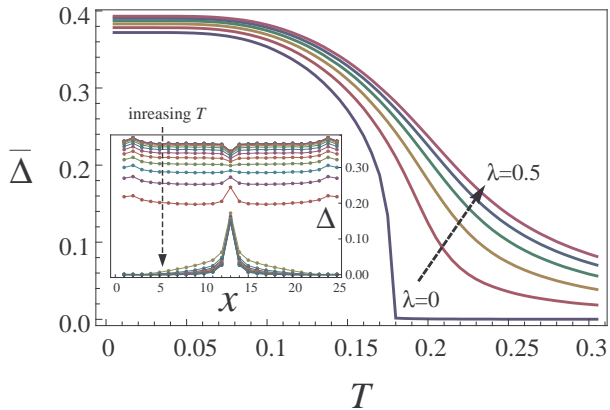}
\caption{ The order parameter, average over cites without a tunneling impurity, as a function of temperature for different values of the impurity tunneling amplitude $\la=0,0.1,...,0.5$. An increase in $T_c$ and a smearing of the transition are clearly seen (see text for numerical parameters). Inset: the real-space dependence of the order parameter around a tunneling impurity, exhibiting a proximity-effect. }\label{Single_Impurity}
\end{figure}

\sec{Summary and Discussion}
In this work we studied the properties of an exciton condensate in the presence of an impurity which induces local tunneling between the layers. It was shown that the impurities induce sub-gap single-particle bound states (Eq.~\ref{Imp-energy}). It is worth pointing out that for strong tunneling the impurity states cross the Fermi level, and a phase-transition occurs, since the ground-state will now have an occupied single-particle fermionic state in it, in similarity to strong magnetic scattering in superconductors \cite{Sasha2}. This transition should have clear spectroscopic features, since the allowed transitions between the bands and the impurity levels, as well as the transitions between the two impurity levels themselves, will depend on their occupation.

In addition, it was demonstrated that around the impurity the condensate order parameter is enhanced (Fig.~\ref{Single_Impurity}). This is a unique situation, and to see this it is useful to compare our system to a superconductor with a magnetic impurity and with a negative-U impurity. In the first case, a single-particle bound-state is formed, but that state disrupts the order parameter in its vicinity, since it acts as a pair-breaker. In the second case the order parameter is enhanced, but there is no single-particle bound state. The tunneling impurity in bilayers combines both these effects. This is due to the unique order parameter of the EC, which corresponds to the interlayer tunneling amplitude.

In the case where the impurity concentration is very large, it is well established that the EC long-range coherence would vanishes due to fluctuations \cite{Shevchenko}. The detailed manner at which the long-range coherence vanishes with increasing impurity concentration is beyond the mean-field level of arguments presented here, and requires calculations in the presence of the order-parameter phase fluctuations (i.e. Kostelitz-Thouless phase fluctuations and the presence of supercurrents). There is preliminary indication that for a small impurity concentration, the supercurrents simply avoid the impurity \cite{Jung-Jung}. How exactly they behave in the presence of a large impurity concentration is left for future studies.

The authors acknowledge valuable discussions with J.-J. Su, M. Lilly and J. Zaanen.  This work was supported by LDRD and  in part, at the Center for Integrated Nanotechnologies, a U.S. Department of Energy, Office of Basic Energy Sciences user facility, by grant No. DE-AC52-06NA25396.

 \end{document}